\documentclass[draft]{agujournal2019}
\usepackage{url} 
\usepackage[inline]{trackchanges} 
\usepackage{soul}
\usepackage{amsmath}

\draftfalse

\journalname{Geophysical Research Letters}

\begin{document}

%
%


\title{Bound on forecasting skill for models of North Atlantic tropical cyclone counts}

%
%




\authors{Daniel Wesley\affil{1}, Michael E. Mann\affil{2}, Bhuvnesh Jain\affil{1}, \\
Colin R. Twomey\affil{3} and Shannon Christiansen\affil{2}}


\affiliation{1}{Department of Physics and Astronomy, University of Pennsylvania, Philadelphia PA 19104}
\affiliation{2}{Department of Earth and Environmental Science, University of Pennsylvania, Philadelphia, PA 19104}
\affiliation{3}{Data Driven Discovery Initiative, University of Pennsylvania, Philadelphia, PA, USA}




\correspondingauthor{Daniel Wesley}{dhwesley@alumni.princeton.edu}

\begin{abstract}
Annual North Atlantic tropical cyclone (TC) counts are frequently modeled as a Poisson process with a state-dependent rate. 
We provide a lower bound on the forecasting error of this class of models. 
Remarkably we find that this bound is already saturated by a simple linear model that explains roughly 50\% of the annual variance using three climate indices: El Ni\~no/Southern Oscillation (ENSO), average sea surface temperature (SST) in the main development region (MDR) of the North Atlantic and the North Atlantic oscillation (NAO) atmospheric circulation index \cite{Kozar2012}. 
As expected under the bound, increased model complexity does not help: we demonstrate that allowing for quadratic and interaction terms, or using an Elastic Net to forecast TC counts using global SST maps, produces no detectable 
increase in skill.  
We provide evidence that observed TC counts are consistent with a Poisson process, limiting possible improvements in TC
modeling by relaxing the Poisson assumption.
\end{abstract}

\section*{Plain Language Summary}
Long range forecasts of TC activity attempt to predict the total number of TC each year well before the season begins. 
These
models often assume the TC count is influenced by climate indicies but otherwise Poisson distributed.  We show that
the error in these forecasts have a lower bound, which current models already achieve.  We show that observed TC
counts are consistent with the Poisson distribution, so our results indicate current models represent the lowest possible
error.  We provide some additional evidence for the bound in two ways.  First we combine climate indices to express more
nuanced influences in the model. Secondly we develop a technique that can use SST to
directly forecast TC, which does not depend on hand-crafting and identifying appropriate climate indices. As predicted
by the bound, neither approach improves forecasting error.  Our results limit possibilities for improving
pre-season forecasts of TC activity.


%
%

%


%
%
%
%

\section{Introduction}
Potential climate influences on the variation over time in North Atlantic Tropical Cyclones (TC) has been a topic of active research for some time.  Numerous prior studies have examined the importance of various climate factors in influencing year-to-year variation in the season number of named storms (annual TC counts). A small number of climate variables have emerged as being particularly important in modeling annual TC counts. It is well known that the El Ni\~no/Southern Oscillation (ENSO) influences seasonal TC activity through its impact on vertical wind shear \cite{Gray1984}, with TC counts being enhanced during El Ni\~no and suppressed during La Ni\~na periods.  Warmer ocean surface temperatures promote the 
formation and development of TCs \cite{Gray1968, Emanuel1995} and numerous studies have thus thus incorporated the impact of sea surface temperatures (SST) over the Main Development Region (MDR; 6$^{\circ}$-18$^{\circ}$N, 20$^{\circ}$-60$^{\circ}$W) during the peak months of the hurricane season (August-October). 
\cite{Hoyos2006, Emanuel2005, Sabbatelli2007, Mann2007}. The North Atlantic Oscillation (NAO) is also relevant to modeling Atlantic TC activity
\cite{Elser2000a, Elser2000b, Elser2003, Elser2006, Mann2007} through its impact on the tracking of storms, which determines in part whether they encounter conditions favorable for tropical cyclogenesis\cite{Elser2000a, Elser2003, Kossin2010}. For recent reviews of the models and methods used for TC forecasting see \cite{Klotzbach2019, Takaya2023} and references therein. 

Previous research has demonstrated that basin-wide TC counts can be effectively modeled as a Poisson process conditioned on key climate state variables \cite{Sabbatelli2007,Kozar2012}. In particular, Kozar et al.\ used forward stepwise Poisson regression to show that the most skillful models for annual TC counts include ENSO, MDR SST, and NAO indices as predictors.

In this study we revisit the Poisson regression framework and show that there is a lower
bound on the forecast error that can be achieved.  The bound is statistical in nature and
applies to Poisson regression models independently of the particular feature set used.
After deriving the bound, we present ``evidence" by considering two extensions of the 
Kozar et al. model.  One approach uses an Elastic Net to forecasts seasonal TC
counts by identifying the most important ``pixels" (2$^{\circ}$ global grid cells) 
in the SST maps.  We also extend the original Kozar et al. model by including nonlinear
``interaction" terms of the features considered in that paper.  The original Kozar et al.
model saturates the bound and neither extension improves it, which is consistent with the predictions of the bound.

In Section \ref{s:poisson} we review the Poisson
regression framework and derive the limit on forecasting skill.  We then introduce
our data sources in Section \ref{s:sst-data}.  In Section \ref{s:methods} we describe the cross-validation and 
Elastic Net methodology.  The results of applying these methods on the data are given in Section 
\ref{s:results} and we conclude in Section \ref{s:conclusions}.

\section{Poisson Regression}\label{s:poisson}

\subsection{Review of Poisson Regression}

Poisson regression has been used in many prior studies of Atlantic TC counts \cite{Elser2000b, Elser2003, Sabbatelli2007, Mann2007, Kozar2012}.  This approach assumes that the probability of observing
a number of TCs $y_t$  in year $t$ is governed by the Poisson distribution
\begin{equation}\label{e:poisson}
P(y_t) = \frac{\lambda_t^{y_t}}{y_t!}e^{-\lambda_t}
\end{equation}
where $\lambda_t$ parameterizes the mean counts expected in year $t$.  
Poisson regression captures variation of observed counts by assuming
$\lambda_t$ varies according to
\begin{equation}\label{e:lambda}
\lambda_t = \exp\left( \beta_0 + \beta_1 x_{1t} + \beta_2 x_{2t} + \cdots 
+ \beta_p x_{pt}\right)
\end{equation}
where $x_{it}$ is the value of ``feature" or ``predictor" $i$ in year $t$, $p$ is the
total number of features, and the $\beta_i$ are 
coefficients which control the influence of feature $i$ on the expected counts.  The coefficient
$\beta_0$, which we sometimes refer to as the ``intercept," controls the mean or unconditional count.  
%
Given a set of observations of TC counts and features $D = \{y_t, x_{it}\}$
we define the log-likelihood function
\begin{equation}\label{e:log-like}
L_{\rm Poisson}(\beta_i|D) = \log P(D|\beta_i)
= \sum_{t=1}^n \left(y_t \sum_j \beta_j x_{jt} - e^{\sum_k \beta_k x_{kt}}\right)
\end{equation}
where terms independent of $\beta_i$ have been dropped.  We then ``fit" the model by choosing
the set of $\hat{\beta}_i$ which maximize $L_{\rm Poisson}$.  
%
%
In this study we quantify forecast quality using mean absolute error (MAE) defined by
\begin{equation}
E = \frac{1}{n} \sum_{t=1}^{n} |y_t - \hat{y}_t|
\end{equation}
where  $\hat{y}_t$ is the model's prediction for the target in observation $t$.  MAE is simply
the expected offset between the forecast and the observed counts.  It is less sensitive to rare 
observations with large residuals $y_t - \hat{y}_t$ than other error measures (such as mean square error)
which promotes statistical efficiency on our relatively small data set.

To 
judge whether a putative feature is truly useful for prediction, or to compare two different sets
of features, we use the forecast errors on validation data computed as part of the N-fold
cross validation procedure (see Section \ref{ss:cross-valid} for a review).
Given two models $M_1$ and $M_2$  we  fit a Poisson regression model and generate the
sequences of forecast errors $\{E_i^{(1)}\}$ and $\{E_i^{(2)}\}$.  This defines the sequence of
differences $\{\Delta_{21,i}\}$ in forecast errors 
\begin{equation}\label{e:Delta}
\Delta_{21,i} = E_i^{(2)} - E_i^{(1)}
\end{equation}
from which we can define the mean change in forecast error and the $t$-statistic
\begin{equation}\label{e:fet}
t_{21} = \frac{\sqrt{N}\,{\rm mean}_i(\Delta_{21,i})}{{\rm std}_i(\Delta_{21,i})}
\end{equation}
The mean change in forecast error indicates whether $M_2$ is an improvement over $M_1$, and 
the $t$-statistic gives us an estimate of the statistical significance of any improvements.
The definition (\ref{e:Delta}) implies that error reduction leads to negative values of
the $t$-statistic, hence throughout this study we follow the convention that negative
$t$-statistics are preferred.

\subsection{Limits on forecasting skill}\label{s:limit}

We show that the broad family
of models which treat TC counts as Poisson distributed face a fundamental limit: 
there is a 
bound on the minimum MAE expected on validation data.  
This bound is saturated by the
explanatory model of \cite{Kozar2012} as well as the Elastic Net described in Sections
\ref{ss:en} and \ref{ss:en-results}.  In this Section we sharpen this surprising claim.

The key insight is that when data is drawn from a Poisson distribution, both the mean
and variance of the observations are controlled by the Poisson rate $\lambda$.  
The observations have nonzero variance, even if $\lambda$ is precisely the correct
Poisson rate of the underlying process.  When we have many draws from a fixed Poisson
distribution with rate $\lambda$, and a forecast $z$ of expected counts,
the expected absolute error on validation data is
\begin{equation}\label{e:mae-lambda}
{\rm err}(z,\lambda) = \sum_{y\ge 0} |y - z| P(y|\lambda)
\end{equation}
where $P(y|\lambda)$ is the probability
of observing $y$ counts with Poisson rate $\lambda$ according to (\ref{e:poisson}).
For fixed $z$ this error has a lower bound $b(z)$ given by
\begin{equation}\label{e:error-bound-1}
b(z) = \underset{\lambda}{\rm min} \; {\rm err}(z,\lambda)
\end{equation}
When we have a forecast $z$ but the true rate $\lambda$ is unknown, the bound $b(z)$
tells us the minimum expected error across all possible values of $\lambda$.

In our application we are given observed counts $y_t$ in each year $t$, but 
the Poisson rates $\lambda_t$ are unobservable.  We can estimate the minimum MAE
on validation data by assuming (1) the distribution of unconditional counts is the same
as our training data and (2) the data is drawn from independent Poisson distributions.
Assumption (1) is satisfied for cross-validated measures of MAE, 
since the cross-validation procedure recycles training data for validation.
In Section \ref{ss:poisson} we test assumption (2) on observed data.

Under our assumptions, the minimum expected MAE is given by averaging 
(\ref{e:error-bound-1})
across all of the training data.  Denoting the minimum expected MAE by $B$ we have
\begin{equation}\label{e:error-bound}
B = \frac{1}{n} \sum_{t=1}^n b(y_t)
\end{equation}
This bound is our estimate for
 the minimum expected error for any model due to drawing
observations from independent Poisson distributions.  It applies to validation 
data only: we can exceed this bound on training data by overfitting.  Also in
any  particular realization of the data, we may exceed this bound by chance.  On
average, or for a large number of observations, we expect it to be accurate.

When we compute $B$ for our training data we obtain $B = 2.51$.  This is consistent
within error bars with the baseline explanatory model performance of
$2.46 \pm 0.10$ given in Table
\ref{t:en-results}.  The bound is also within the error bars of the
EN model $2.51 \pm 0.14$. While the baseline model appears to be slightly ``lucky" at
the $0.5\sigma$ level, both of these models saturate the error bound.

The estimate (\ref{e:error-bound}) passes some nontrivial checks.  The average
TC count in our data is close to 10.  We generated
Monte Carlo ``observations" from an underlying Poisson distribution with $\lambda = 10$
in every year, then computed the bound $B$ using these ``observations."
This gave a minimum MAE estimate within about 2\% of the correct value.  
We have also used the predicted TC counts
from the explanatory model of ref \cite{Kozar2012} in place of the minimization over 
$\lambda$, giving an estimate of 2.57 on our data.  This estimate is higher than our
bound, as it should be: any concrete model should give a bound no lower than our 
theoretical bound.

The bound $B$ defined by (\ref{e:error-bound}) is 
independent of the model used to generate the predictions of TC rates $\lambda_t$ in
each year.  Provided the model hews to the assumption of independent Poisson distributed
observations, the bound will apply.  Given that we already have models that saturate
the bound, more sophisticated modeling techniques
that enhance the $\lambda_t$ prediction are unlikely to improve performance.
Better performance will likely require an approach that does not treat TC counts
as drawn from independent Poisson distributions -- but we show in Section 
\ref{ss:poisson} that the historical TC counts are consistent with the Poisson assumption, which constraints the space for improved modeling. 

We have framed the bound $B$ in terms
of the MAE cost function and Poisson distributions.
We can easily define analogous bounds for different
cost functions and distributions provided we keep the assumption of independent draws
and additive cost functions.  For a cost function $f$ and a distribution parameterized
by parameters $\theta$ we can define
\begin{equation}
{\rm err}(z, \theta) = \sum_{y \geq 0} f(y-z) P(y|\theta)
\end{equation}
and the analogues of (\ref{e:error-bound-1}) and (\ref{e:error-bound}) are
\begin{equation}
b(z) = \underset{\theta}{\rm min} \, {\rm err}(z, \theta)
\end{equation}
and
\begin{equation}
B = \frac{1}{n}\sum_{t=1}^n b(y_t)
\end{equation}
Different error functions will give different numerical bounds.

\section{Data}\label{s:sst-data}

In this paper we use climate indices, global SST data, and annual TC count series. The climate indices used are well described in previous references, eg \cite{Kozar2012}.
The key indices are the in season August-October (ASO) mean temperature in the Main Development Region
(MDR), the December-February (DJF) Nino 3.4 index and the late-to-post-season boreal winter
December-March (DJFM) North Atlantic Oscillation (NAO) index.
For the global SST data we use the NOAA Extended Reconstructed SST v5 (ERSSTv5) \cite{ERSSTv5}.  This data set provides coverage over nearly the entire ocean surface
with mean monthly SST temperatures in a 2$^{\circ}$ x 2$^{\circ}$ 
latitude and longitude grid.
For TC counts, we use the adjusted TC counts published in \cite{VK08}.  This timeseries
corrects for the improvement over time in the detction of TCs from technological advances
such as aircraft reconnaissance and satellites.

\section{Methods}\label{s:methods}

\subsection{Cross Validation}\label{ss:cross-valid}

Throughout this paper we use the standard statistical meta-algorithm of $N$-fold cross-validation (NFCV). 
NFCV requires (1) a partitioning of our data set into $N$ subsets or ``folds," and (2)
a measure of forecast error.  NFCV ensures forecast quality is always evaluated on data that was 
``held out" of training which provides protection against overfitting and a realistic test of the skill of
the model for real forecasting problems.  

NFCV works as follows: for each $i=1,2, ... N$ we construct a training set $T_i$ by 
aggregating all folds except the $i^{th}$ one, and use the $i^{th}$ fold as the validation set $V_i$.
Train the model using only data in $T_i$, make forecasts on 
$V_i$ and compute the forecast error measure $E_i$.
Note that the measure of forecast error may be different from the cost function employed for training
the model.  The sequence $\{E_i\}$ of forecast error measurements can be used to compute absolute
model performance.  When comparing two models, the two error sequences for the models can be used to
assess the statistical significance of any observed out performance of one model relative to the other.
For this study, we divide the 140 yearly observations 1880-2019 into $N=5$ equally sized folds 1880-1907,
1908-1935, etc.  This maximizes the chances that adjacent years (which may not be statistically 
independent) are assigned to the same fold.  

\subsection{Elastic Nets}\label{ss:en}

Previous analyses \cite{Kozar2012} have generally incorporated SST data using features inspired
by an understanding of the processes involved in hurricane formation and climate processes.
Here we investigate whether there is additional information in global SST data that is not
expressed in the existing hand-crafted features.
While we are only using SST data, other
drivers of Atlantic TC counts (such as wind or current patterns) may influence the SST field
and thus be incorporated indirectly into the model. For example, the Nino3.4 index is based on the tropical Pacific SST field, but it is actually a metric of how the ENSO phenomenon impacts wind patterns in the tropical Atlantic that govern TC formation.

We will attempt to 
incorporate the full SST data by designing an algorithm that takes global SST map-level data and uses it to 
forecast the annual TC count directly. 
The technique is designed to ignore temperature observations
that are not useful for forecasting TCs.
Each month, the ERSSTv5 data set provides temperature data on roughly $8.8 \times 10^3$ 
grid points covering the globe, while we have only $n = 142$ observations of annual 
TC counts, so we are deeply in the $p\gg n$ statistical regime.

A given temperature observation $\tau_{xymt}$
in ERSSTv5 is specified by four indices $(x, y, m, t)$
giving the latitude and longitude indices $(x, y)$ of the grid location, 
the observation month $m$ (January-December) and the observation year $t$.  We process the features
into normalized versions by computing the mean and standard deviations over $t$ using 
the training data.
\begin{equation}
\mu_{xym} = {\rm mean}_t(\tau_{xymt}),\;
\sigma_{xym} = {\rm std}_t(\tau_{xymt})
\end{equation}
We then define a consolidated ``pixel" index $i$ together with a bijection $i \leftrightarrow (x,y,m)$.
In our terminology a pixel refers to a specific geographical location together with an observation month.
Using the pixel index $i$ we define normalized features $z_{it}$ by
\begin{equation}\label{e:z-def}
z_{it} = \frac{\tau_{it} - \mu_i}{\sigma_i}
\end{equation}
On the training data $z_{it}$ has zero mean and unit standard deviation by construction.
When performing cross-validation (See Section \ref{ss:cross-valid}) it is crucial that no information
about the validation data ``leak" into the training phase.  On validation data, we use
the same formula (\ref{e:z-def}) but apply the $\mu_i$ and $\sigma_i$ computed from training data.

The algorithm is based on the Elastic Net (EN) \cite{Zou2005} and adapted to the Poisson regression
framework.  To construct the objective function $L_{\rm EN}$ for training 
we use the SST features $z_{it}$ in the Poisson regression log-likelihood
(\ref{e:log-like}) but add additional regularization terms inspired by the Elastic Net
\begin{equation}\label{e:Elastic Net}
L_{\rm EN}(\beta_0, \beta_1, \beta_2, \dots \beta_X, \lambda_1, \lambda_2) = 
L_{\rm Poisson}(\beta_0, \beta_1, \beta_2, \dots \beta_X) - \lambda_1 \sum_{i=1}^X | \beta_i | 
- \frac{1}{2} \lambda_2 \sum_{i=1}^X \beta_i^2
\end{equation}
where $X$ is the total number of pixels and $\lambda_1 \ge 0$ and $\lambda_2 \ge 0$ are 
regularization parameters.  The $\lambda_1$ encourages sparsity and can perform a kind of
variable selection by encouraging weak features to acquire zero coefficients.  The $\lambda_2$
term promotes grouping of highly correlated features and encourages the model to assign 
them similar coefficients. Note that the intercept term $\beta_0$ does not have any penalties
applied to it.

The EN is trained using a two-step process.  In the first step we use a gradient ascent algorithm
to find good values of the coefficients.  Denoting by $\beta_i^s$ the value of coefficient $\beta_i$ in
step $s$ of this process, we initialize $\beta_i^0 = 0$ for $i > 0$ and $\beta_0^0 = \log {\rm mean}_t y_t$.
Then we adjust the values of the $\beta_i^s$ by
\begin{equation}
\beta_i^{s+1} = \beta_i^s + \alpha \frac{\partial L_{\rm EN}}{\partial \beta_i}
\end{equation}
where $\alpha$ is the learning rate.  We have found $\alpha = 10^{-6}$ works well in practice.
Once gradient ascent ceases to improve the objective function $L_{\rm EN}$ on training data we
halt the algorithm and assign preliminary values $\beta_i' = \beta_i^{s'}$ where $s'$ is the
step with the maximum value of the objective function.

In the second training step, we adjust values of the coefficients.  The EN penalties encourage
coefficients to shrink toward zero and hence forecasts of TC counts will be biased.  To reduce
this bias, we construct an aggregated SST feature $w_t$ using coefficients from the first stage
fits
\begin{equation}
w_t = \sum_{i=1}^X \beta_i' z_{it}
\end{equation}
we then do a second Poisson regression on the training data using the two features $\{\beta_0, w_t\}$.  This second Poisson regression yields coefficients $\gamma_0$ on the
constant term and $\gamma_{\rm SST}$ on the aggregate SST feature.  To generate forecasts
we use the Poisson rate
\begin{equation}
\lambda_t = \exp \left(
\gamma_0 \beta_0' + \gamma_{\rm agg} w_t \right)
\; {\rm with} \; w_t = \sum_{i=1}^X \beta_i' z_{it}
\end{equation}
This is completely equivalent to standard Poisson forecasting 
using the intercept and SST pixel temperatures $z_{it}$
with coefficients $\beta_0 = \gamma_0\beta_0'$ and $\beta_i = \gamma_{\rm agg}\beta_i$ for $i>0$.
However the EN procedure assigns different values to the coefficients than the Poisson regression
procedure.

\section{Results}\label{s:results}

Throughout the results section, we will use the model of ref \cite{Kozar2012} as our ``baseline"
model.   We distinguish two versions of the model.  The key difference between the model is the
month range for MDR temperatures.   The ``explanatory" version is the one described
in ref. \cite{Kozar2012}, using the in-season (ASO) MDR mean temperature.  This version of the model
is useful for understanding linkages between climate variables and the TC count.  However, because
the ASO MDR temperature is not known until the end of the hurricaine season, this model cannot be used
to predict TCs before the season commences.  We define a ``predictive" version of the model which 
uses the April (P) MDR mean temperature.  This version of the model can be used to predict the 
seasonal TC count before the season begins.

\subsection{Consistency with Poisson assumption}\label{ss:poisson}

The argument presented in Section \ref{s:limit} depends crucially
on observations being generated by independent Poisson draws in each observation year, so it
is important to check this assumption on data.  Ref \cite{Sabbatelli2007} used a chi-squared
statistic which measures consistency with Poisson-distributed observations.  The
test statistic is
\begin{equation}
x = \sum_{t=1}^n \frac{(y_t - \hat{y}_t)^2}{\hat{y}_t}
\end{equation}
where $y_t$ are the observed counts and $\hat{y}_t$ are the counts forecasted by the 
model in year $t$.  We fit the model of \cite{Sabbatelli2007, Kozar2012} over our 
entire sample (no cross validation) and used its 
predictions as $\hat{y}_t$ to compute $x$, which gives a result fully consistent with
Poisson distributed residuals ($p=0.84$).  The serial autocorrelation of residuals
to this model is 11\%, mildly positive but consistent with zero ($z = 1.3$).  Both of
these measurements support our assumption of the observed TC counts being generated
by independent Poisson distributions each year.  Hence we expect the limit described in Section 
\ref{s:limit} to hold in this study.

\subsection{Elastic Net}\label{ss:en-results}

We find that the EN is able to produce comparable performance to the 
``predictive" baseline model
of Ref. \cite{Kozar2012}, despite the fact that it uses SST data only.  
The predictive baseline model uses features derived from prior
year December through present year April observables.  To keep the comparison consistent,
we provide the EN with SST data from prior December through the current April as well.
There is a broad region in the $(\lambda_1, \lambda_2)$ plane that gives good EN performance.
With the choice\footnote{Using a coarse grid search, the minimum error we have found 
uses $\lambda_1 = 0.577$, $\lambda_2 = 3.33 \times 10^3$ and gives cross-validated forecast
errors of $2.73 \pm 0.11$.}
$\lambda_1 = 1$, $\lambda_2 = 10^4$ the EN achieves a cross-validated forecast
error of $2.78 \pm 0.12$, to be compared with $2.81 \pm 0.24$ for the baseline model.
(See Table \ref{t:en-results})
While the mean error is slightly lower than the baseline model, the statistical significance is
low ($t = -0.12$).  The error is dominated by the variation in baseline model forecast errors:
forecast errors for the EN vary by about half as much across validation folds.

\begin{table}
\begin{center}
  \begin{tabular}{| l | c | c | c | c | c | }
\hline
model               & \multicolumn{2}{c|}{absolute} & \multicolumn{3}{c|}{relative} \\ 
                    &  error & std  & error & std  & $t$-stat \\
\hline
\hline
baseline predictive  & 2.81 & 0.24 &       &       &      \\ 
EN $(1, 10^4)$       & 2.78 & 0.12 & -0.03 &  0.21 & -0.12 \\
\hline
\hline
baseline explanatory & 2.46 & 0.10 &       &       &      \\ 
EN $(0.316, 10^4)$     & 2.51 & 0.14 & 0.05 &  0.12 &  0.41 \\ 
\hline
  \end{tabular}
\end{center}
\caption{Model performance metrics for baseline and EN models, in predictive
mode (top) and explanatory mode (bottom).  We report cross-validated 
forecast errors in absolute terms and relative to the 
corresponding baseline model as defined in (\ref{e:fet})}
\label{t:en-results}
\end{table}

\begin{figure}
\centering
\includegraphics[width=6in]{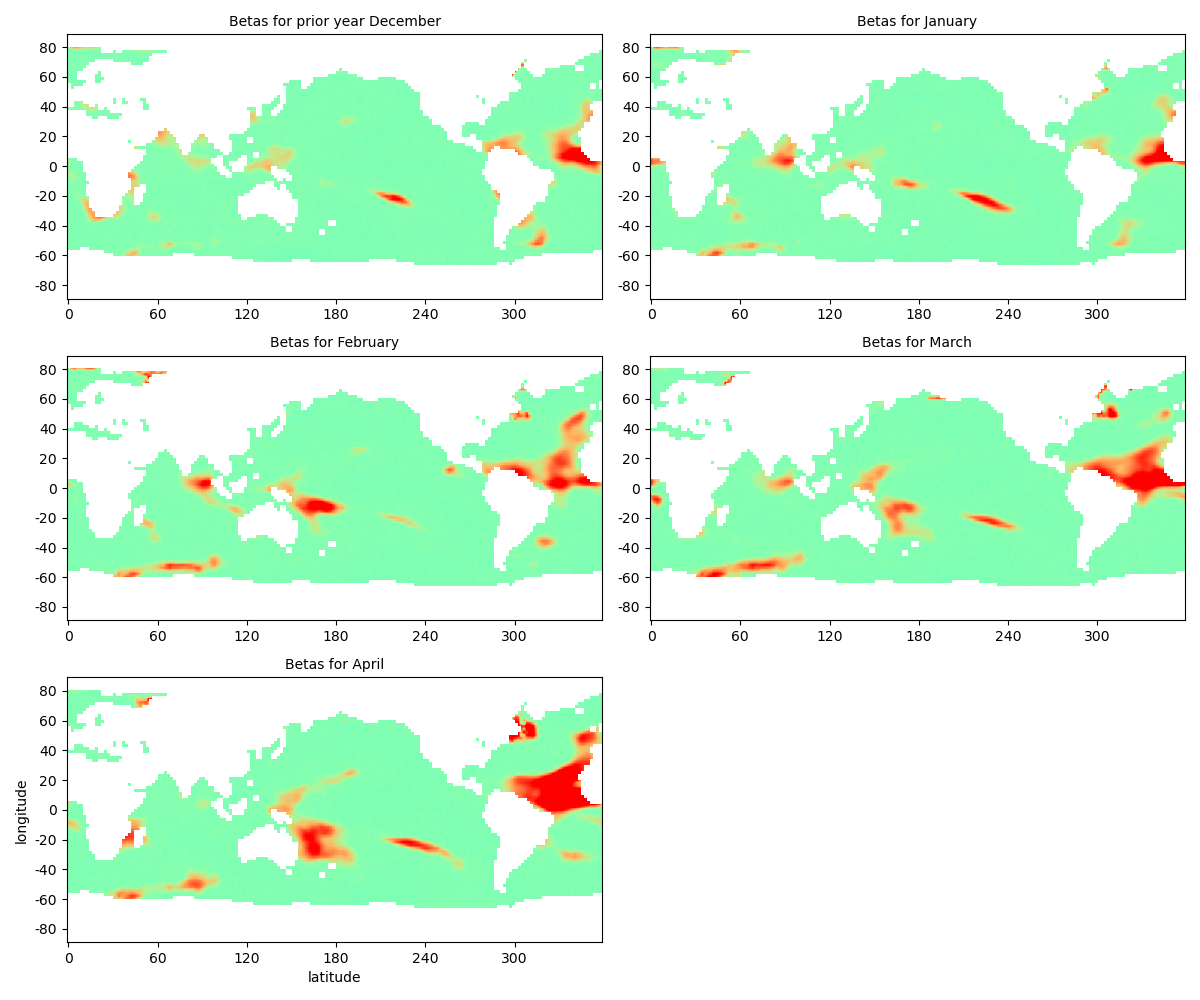}
\caption{This figure shows the coefficients $\beta_i$ for the EN fit with $\lambda_1 = 1$, 
$\lambda_2 =10^4$.  Each feature $i$ corresponds to a specific grid location and month of
the year, so $\beta_i$ measures the weight the model places on each observation.}
\label{f:en-map}
\end{figure}

In addition to providing a forecast, the EN also generates a map showing what oceanic regions
and months provide SST information relevant to forecasting TCs.  Each feature $i$ in the
EN corresponds to SST at a specific grid location during a specific month of the year.  Hence
the coefficients $\beta_i$ give a measure of the weight that the EN puts on that location and
month in generating its forecasts.  
The $\beta_i$ for the fiducial EN model are illustrated in Figure \ref{f:en-map}.  Most of the
pixels are close to zero, reflecting the influence of the $\lambda_1$ penalty in eliminating
some features.  The maps clearly show the weight placed on the area near the MDR, with the
weight increasing as the year progresses.  This reflects the importance of the MDR SST in 
predicting TC counts.  We also see some features in the Pacific.  It seems plausible that this
is related to ENSO.

Results are similar when we construct an EN by analogy to the ``explanatory" baseline model.
The explanatory baseline uses data through October, so to keep the comparison consistent
the EN uses data from prior year December through current October.  We find that the EN
with $\lambda_1 = 0.316$ and $\lambda_2 = 10^4$ 
achieves cross-validated forecast error of $2.51 \pm 0.14$ as compared with 
$2.46 \pm 0.10$ for the baseline model.  The performance difference is not statistically 
significant ($t=0.41$) so we see that comparable performance to the baseline model is
achieved. See Table \ref{t:en-results} for details.

\begin{figure}
\centering
\includegraphics[width=6in]{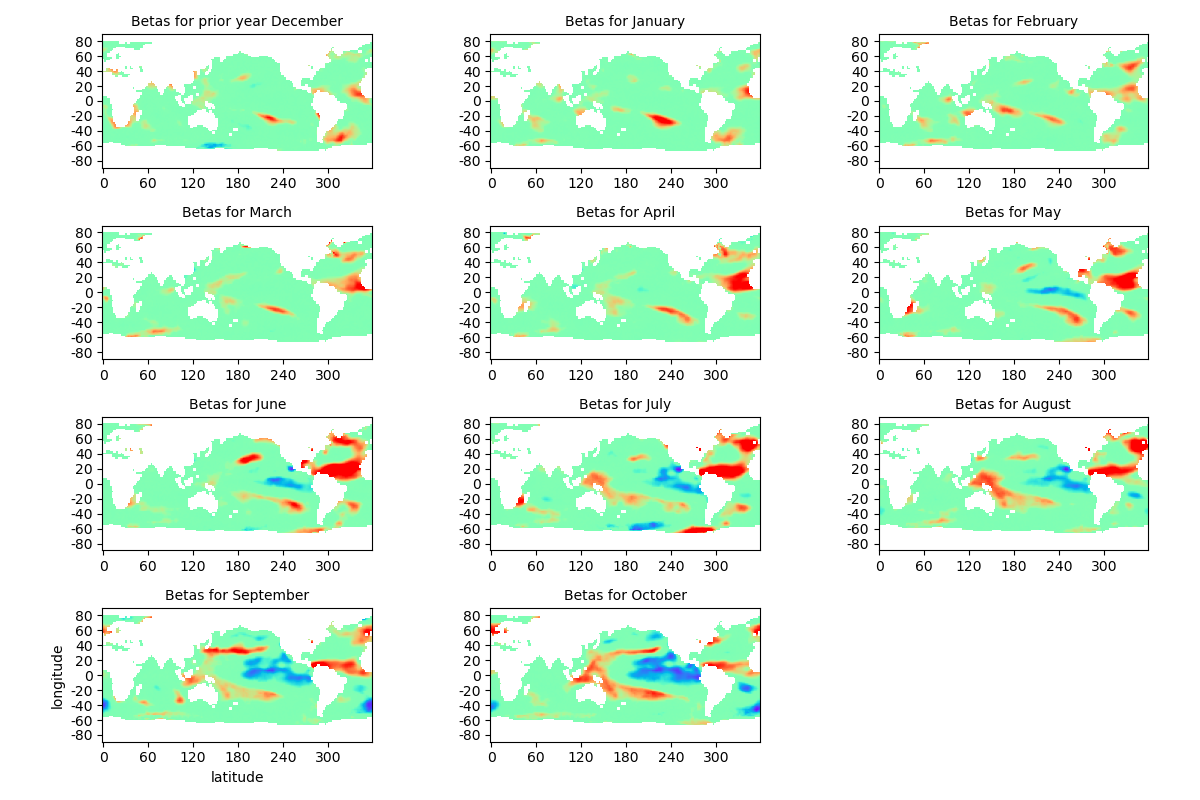}
\caption{This figure shows the coefficients $\beta_i$ for the EN fit with $\lambda_1 = 0.316$, 
$\lambda_2 =10^4$ in explanatory mode.  Each feature $i$ corresponds to a specific grid location and month of the year, so $\beta_i$ measures the weight the model places on each observation.}
\label{f:en-map-explanatory}
\end{figure}

The maps for the explanatory EN are shown in Figure \ref{f:en-map-explanatory}.  In this 
figure we see a similar pattern to the predictive EN seen in Figure \ref{f:en-map} with the 
MDR region acquiring a prominent positive weight.  For later months the El Ni\~no region
acquires a prominent negative weight.  This is consistent with the baseline model, in which
higher temperatures in the Ni\~no 3.4 region correspond to a decrease in the TC prediction.


\subsection{Nonlinear interactions}

The Poisson regression framework allows us to explore nonlinear interactions between the 
predictors in the baseline model.  We do this by constructing ``product features" and testing
whether they improve forecasting errors when added to the baseline model.

Given two features $x$ and $y$, we define the product feature $x \star y$ by demeaning $x$ and
$y$ on the training data, then multiplying the resulting values together observation by
observation.  On validation data, we use the same means derived from the training data, so this
procedure is consistent with the cross-validation procedure.  We then test a series of models,
one per product feature, obtained by adding the product feature to the features in the
baseline model.

Table \ref{t:nonlinear} summarizes the result of this test for the ``explanatory" baseline
model.  Of the candidate predictors tested, only  the product \texttt{nino34\_djf * nao\_djfm}
 showed a reduction in forecast error
when added to the baseline model: forecast error decreases by a small amount ($0.03$)
but the statistical significance of the effect is fairly strong ($t=-4.71$) and other 
checks show it is quite consistent across cross-validation folds.

\begin{table}
\begin{center}
  \begin{tabular}{| l | c | c | c | c | c | }
\hline
model               & \multicolumn{2}{c|}{absolute} & \multicolumn{3}{c|}{relative} \\ 
                    &  error & std  & error & std  & $t$-stat \\
\hline
\hline
baseline explanatory & 2.46 & 0.10 &       &       &      \\ 
\texttt{mdr\_aso * mdr\_aso      } & 2.63 & 0.19 & 0.16 & 0.126 &  1.30 \\ 
\texttt{mdr\_aso * nino34\_djf   } & 2.50 & 0.09 & 0.04 & 0.009 &  4.36 \\ 
\texttt{mdr\_aso * nao\_djfm     } & 2.47 & 0.10 & 0.00 & 0.009 &  0.37 \\ 
\texttt{nino34\_djf * nino34\_djf} & 2.47 & 0.11 & 0.01 & 0.013 &  0.84 \\ 
\texttt{nino34\_djf * nao\_djfm  } & 2.43 & 0.11 & -0.03 & 0.007 & -4.71 \\ 
\texttt{nao\_djfm * nao\_djfm    } & 2.47 & 0.10 & 0.00 & 0.007 &  0.74 \\ 
\hline
\hline
baseline predictive & 2.81 & 0.24 &       &       &      \\ 
\texttt{mdr\_p * mdr\_p          } & 2.88 & 0.27 & 0.08 & 0.042 &  1.85 \\ 
\texttt{mdr\_p * nino34\_djf     } & 2.94 & 0.31 & 0.13 & 0.103 &  1.28 \\ 
\texttt{mdr\_p * nao\_djfm       } & 2.83 & 0.25 & 0.02 & 0.012 &  1.63 \\ 
\texttt{nino34\_djf * nino34\_djf} & 2.85 & 0.25 & 0.04 & 0.012 &  3.14 \\ 
\texttt{nino34\_djf * nao\_djfm  } & 2.81 & 0.24 & 0.00 & 0.020 &  0.09 \\ 
\texttt{nao\_djfm * nao\_djfm    } & 2.81 & 0.23 & 0.01 & 0.011 &  0.69 \\
\hline
 \end{tabular}
\end{center}
\caption{Model performance metrics with candidate
nonlinear terms added, in predictive
mode (top) and explanatory mode (bottom).  We report cross-validated 
forecast errors in absolute terms and relative to the 
corresponding baseline model as defined in (\ref{e:fet})}
\label{t:nonlinear}
\end{table}

Unfortunately when carrying out the same test using the features in the predictive model we do not
see a similar reduction in error from adding the product feature
\texttt{nino34\_djf * nao\_djfm}.


In addition to the features found in the baseline model, we have also examined all product
features using the full 10 predictor set studied in \cite{Kozar2012}.  This study revealed no
interesting product features, except those trivially related to the 
\texttt{nino34\_djf * nao\_djfm} one described above.  We have also explored some other forms
of nonlinear interaction that did not reveal additional features of interest.

\section{Conclusions}\label{s:conclusions}

Prior work modeling annual TC counts as a Poisson process with a state-dependent rate has revealed that roughly 50\% of the annual variance can be predicted using three climate indices: El Ni\~no/Southern Oscillation (ENSO), average SST in the MDR of the North Atlantic and  North Atlantic oscillation (NAO) atmospheric circulation index \cite{Kozar2012}.
Here, we have explored the limits of forecast accuracy in models of this type.
In this work, we presented an argument that any model that treats observed TC counts as draws from a Poisson distribution must have a lower bound on the cross-validated forecast error, and that the model of Ref \cite{Kozar2012} saturates this bound.

We also show that, as expected under the bound, additional model complexity does not help.
Using Atlantic tropical cyclone (TC) data over 1878-2020 and carefully cross-validating 
we have demonstrated that an Elastic Net (EN) model based on global sea surface temperature (SST) maps can at most produce comparable performance to the models using climate indices.
Extending the Ref \cite{Kozar2012} feature set through nonlinear features does not improve performance.
Lastly, we validate that the residual variance and autocorrelation for these models are indeed consistent with Poisson-distributed TC counts.  Hence we conclude these models realize the ``best'' possible performance achievable when annual TC counts are modeled as independent Poisson draws.

To achieve better performance, a model would need to relax one of the underlying assumptions
in the bound: namely, independent draws from Poisson distributions each year.  This presents
a challenge since we have explicitly tested these assumptions and all results so far are
consistent with them.  However, one could imagine a subtle relationship in counts across different years which is not captured by our tests, perhaps modulated by a conditioning variable we have not yet identified.  This would violate the independence assumption and change the nature of the
bound. We hope future work will illuminate this issue further.

Finally, we note that our study shares some limitations with all attempts at modeling TC counts, namely that the available historical record represents a small number of observations and that possible effects of climate change may lead to a change in the causal relationships that are difficult to discern from recent data. 

\section*{Open Research Section}

The adjusted TC counts published in \cite{VK08} and climate indices are publicly available from the Penn State/IBM
Nittany AI Alliance which can be accessed on GitHub at https://github.com/NittanyAiAlliance/IBM-Weather/tree/main/Huriccane\_Data.  The 
NOAA Extended Reconstructed SST V5 data (ERSSTv5) \cite{ERSSTv5} is provided by the NOAA PSL, Boulder, Colorado, USA
with access information at https://psl.noaa.gov/data/gridded/data.noaa.ersst.v5.html.  
The figures for this manuscript were generated using the Anaconda software 
distribution version 24.7.1 available from https://www.anaconda.com.  Analysis and figure generating 
code, as well as copies of the adjusted TC counts and climate index data is publicly available on GitHub at 
https://github.com/wes137/bounds\_tc\_counts \cite{OurCode}.

\acknowledgments
We thanks Gary Bernstein and Mike Jarvis for helpful discussions and the Data Driven Discovery Initiative at Penn's School of Arts and Sciences for partial support.

%
%

\bibliography{article}

%
%
%
%
%

\end{document}